%% file: eprint_dpf2013_bodek.tex
\documentclass[12pt]{article}
\usepackage{graphicx}

\newcommand\pubnumber{DPF2013-??}
\newcommand\pubdate{\today}

\def\rochester{$^a$Department of Physics and Astronomy\\
University of Rochester, Rochester, NY 14627 USA\\
$^b$Hampton University; Hampton, Virginia, 23668 USA}
\def\support{\footnote{Work supported by the US Department of Energy.}}

\def\Title#1{\begin{center} {\Large #1 } \end{center}}
\def\Author#1{\begin{center}{ \sc #1} \end{center}}
\def\Address#1{\begin{center}{ \it #1} \end{center}}

\newcommand\pubblock{\rightline{\begin{tabular}{l} \pubnumber\\
         \pubdate  \end{tabular}}}
\newenvironment{Abstract}{\begin{quotation}  }{\end{quotation}}
\newenvironment{Presented}{\begin{quotation} \begin{center} 
             PRESENTED AT\end{center}\bigskip 
      \begin{center}\begin{large}}{\end{large}\end{center} \end{quotation}}

\input econfmacros.tex

\begin{document}
\begin{titlepage}
\pubblock

\vfill
\Title{Further Studies of Transverse Enhancement in Quasielastic Electron Scattering}
\vfill
\Author{ A. Bodek$^{a,}$\support, H.S. Budd$^{a}$, M. E. Christy$^b$, T. N. S. Gautam$^b$}
\Address{\rochester}
\vfill
\begin{Abstract}
In a previous communication we  reported on  a  parametrization of  the observed  enhancement in the transverse electron  quasielastic (QE) response function for nucleons bound in carbon as a function of the square of the four momentum transfer ($Q^2$) in terms of a  correction to the magnetic form factors of bound nucleons.   That parametrization was used to predict the overall magnitude and $Q^2$ dependence 
of the cross section for  neutrino quasielastic scattering on nuclear targets.  In this paper, we extend the study to include parametrizations of both the $Q^2$ as well as  the energy transfer ($\nu$) dependence of the transverse enhancement. These parametrization can be used
to give a more complete two dimensional description of  the neutrino quasielastic scattering process on nuclear targets,  which is essential for precision studies of mass splitings and mixing angles in neutrino oscillation experiments.
\end{Abstract}
\vfill
\begin{Presented}
DPF 2013\\
The Meeting of the American Physical Society\\
Division of Particles and Fields\\
Santa Cruz, California, August 13--17, 2013\\
by Arie Bodek\\
\end{Presented}
\vfill
\end{titlepage}
\def\thefootnote{\fnsymbol{footnote}}
\setcounter{footnote}{0}

\section{Introduction}

In a previous communication\cite{firstpaper}  we  reported on  a  parametrization of  the observed  enhancement in the transverse electron  quasielastic (QE) response function for nucleons bound in carbon as a function of the square of the four momentum transfer ($Q^2$) in terms of a  correction to the magnetic form factors of bound nucleons. In this paper, we provide additional details about both
the  of the  $Q^2$ as well as  the energy transfer ($\nu$) dependence of the transverse enhancement
\section{The method}

In our previous paper,  we used a fit to existing electron scattering differential cross sections  on nuclei, including preliminary data from the JUPITER 
collaboration\cite{JUPITER}  (Jefferson lab experiment E04-001).  The fit \cite{vahe-thesis} (developed  by P. Bosted and V. Mamyan) provides a description of inclusive electron scattering differential cross sections on a range of nuclei with $A > 2$.  It is an extension of fits to the 
free proton~\cite{resp} and deuteron~\cite{resd} cross sections. The fit is utilized for calculations  of the radiative corrections for the JUPITER analysis~\cite{vahe-thesis}. Experiment   
E04-001 was designed to measure separated  longitudinal and transverse 
structure functions for a range of nuclear targets.  The measurements of 
the vector structure functions cover both the  quasi-elastic and resonance 
regions.  

 A brief description 
of the fit is given in \cite{vahe-thesis}. Additional details are presented in our previous paper\cite{first paper}
The inclusive fit is a sum of four components: 
\begin{itemize}
\item The longitudinal QE contribution calculated for independent nucleons (smeared by Fermi motion in carbon)
\item  The transverse  QE contribution calculated for independent nucleons (smeared by Fermi motion in carbon)
\item  The contribution of inelastic pion production processes (smeared by Fermi motion in carbon).
\item  A transverse excess (TE) contribution (determined by the fit). The transverse enhancement contribution is a modified Gaussian, with amplitude, width and peak position which are allowed to vary in the fit.
\end{itemize}
The QE model used in the Bosted-Mamyan fit is the super-scaling model\cite{super} of  Sick, Donnelly, and  Maieron.  In this paper we report on an updated analysis in which we have modified the functional form for the transverse enhancement to better fit the electron scattering data. 

\begin{figure}[htb]
\centering
\includegraphics[height=3.5in, width=6.1in]{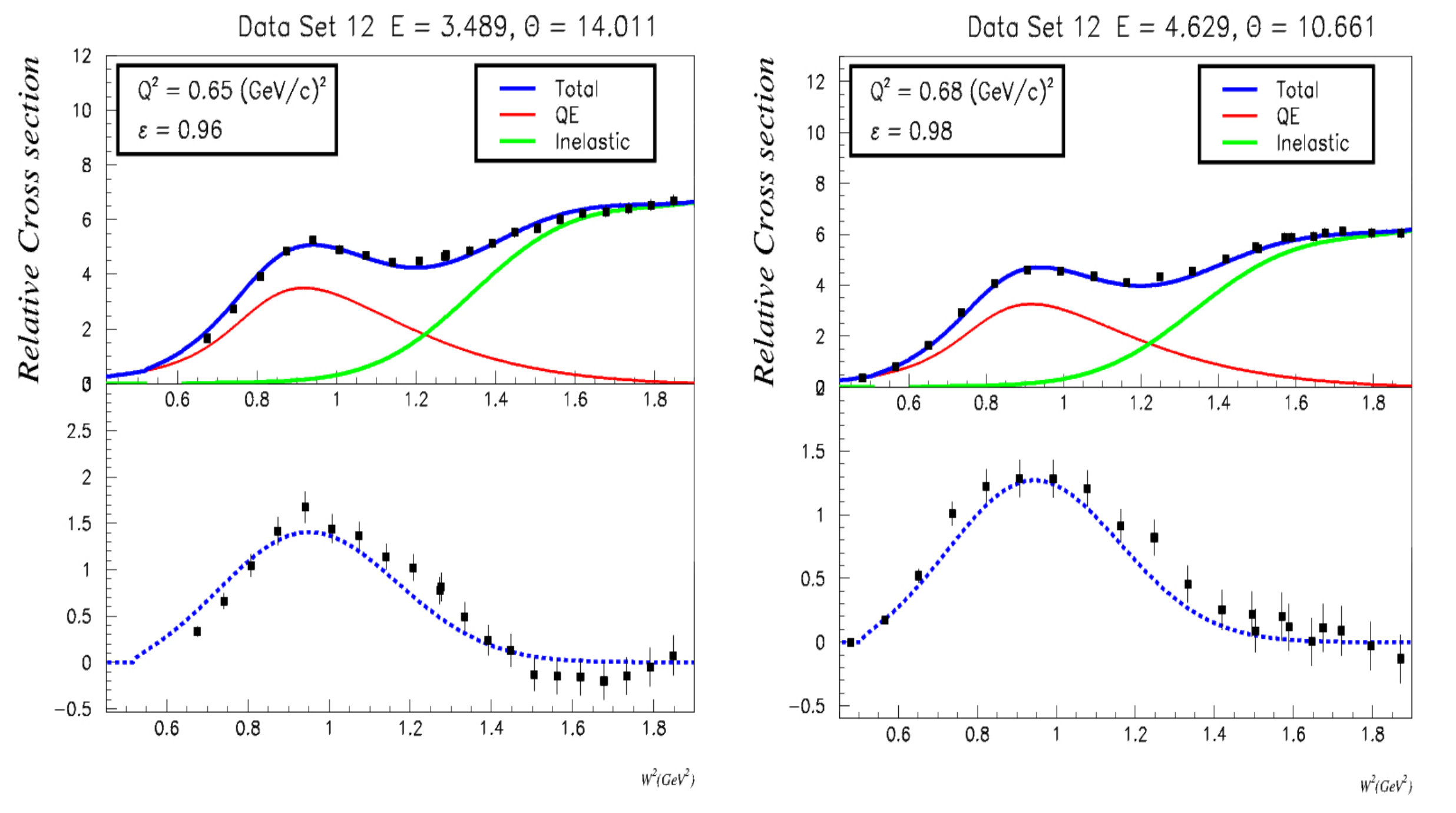}
\caption{Examples of two  fits to preliminary electron scattering differential cross sections for  carbon target. The top panels shows the contributions from the  independent nucleon quasi elastic process (in red) as a function of the final state invariant mass $W^2$. The contribution from inelastic pion production processes is shown in green.  The bottom panels shows the residual of the data minus the sum of QE and pion production processes.  The fit to the residual, which is the transverse excess (TE) contribution, is shown a the dashed blue line.}
\label{example}
\end{figure}

Figure~\ref{example} shows examples of two  fits to preliminary electron scattering differential cross  sections for a carbon target. The top panels show the contributions from the  independent nucleon quasi elastic process (in red) as a function of the final state invariant mass $W^2$. The contribution from inelastic pion production processes is shown in green.  The bottom panels shows the residual of the data minus the sum of QE and pion production processes.  The fit to the residual, which is the transverse excess (TE) contribution, is shown a the dashed blue line.

We extract the integrated transverse enhancement ratio as a function of $Q^2$ by integrating the various contributions to the fit up to $W^2=1.5~ GeV^2$. Here
 $${\cal R}_{T} =\frac {QE_{transverse}+TE}{QE_{transverse}}$$
 We also extract the peak positions (in $\nu$)  and RMS width (in $\nu$) of both the QE and transverse enhancement contributions. 

\begin{figure}[htb]
\centering
\includegraphics[height=3.5in, width=6.1in]{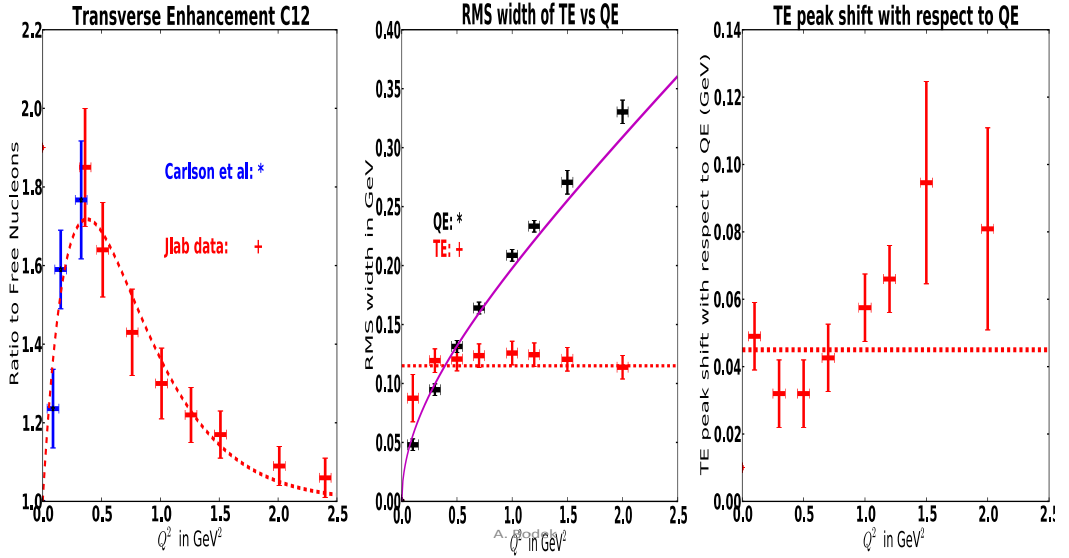}
\caption{ Left panel: The ratio  ($R_T$)of the integrated transverse  quasielastic response function for Carbon 12
 to the integrated transverse response function for the independent nucleon model as a function
 of $Q^2$.  Middle panel: The RMS width (in $\nu$) for the  transverse enhancement 
component as compared to the RMS with of  quasielastic scattering from independent nucleons as a function of $Q^2$.  Right panel: The shift in the peak position  (in $\nu$) of the transverse enhancement with respect to the 
peak position for quasielastic scattering from independent nucleons as a function of $Q^2$.}
\label{TEvsQ2}.
\end{figure}

\section{Results}

 The left panel of Fig. \ref{TEvsQ2} shows the ratio  ($R_T$)of the integrated transverse  quasielastic response function for Carbon 12
 to the integrated transverse response function for the independent nucleon model as a function
 of $Q^2$. The ratio can be parametrized as 
 $R_T$= 
 1 + AQ$^2$ e$^{-Q^2/B}$, 
 with  A=5.19 GeV$^{-2}$ and B=0.376 GeV$^2$. 
 
The middle panel of   Fig. \ref{TEvsQ2} shows the RMS width (in $\nu$) for the  transverse enhancement 
component as compared to the RMS with of  quasielastic scattering from independent nucleons as a function of $Q^2$. The RMS width of the TE contribution is about 0.115 GeV. This is to be  compared to the RMS width for QE scattering from nucleons which (as expected) increases with $Q^2$.
The RMS width for QE scattering from independent nucleons (in $\nu$) can be represented by the function  $RMS_{QE}$= C$q_3$  Here , the RMS width is in GeV,    C=0.174 and  $q_3=\sqrt{Q^2(1+Q^2/4M^2)}$ is the magnitude of the 3-momentum transfer to the nucleon in GeV.

The right panel of Fig. \ref{TEvsQ2}  shows the shift in the peak position  (in $\nu$) of the transverse enhancement with respect to the peak position for QE scattering from independent nucleons as a function of $Q^2$ . The
shift in the  peak position of the transverse enhancement contribution appears to be independent of $Q^2$. On average,  the tE contribution is shifted by 0.045 GeV  towards  higher $\nu$.  This is expected if  the TE contribution  originates from a two nucleon process.


\end{document}

%% file: econfmacros.tex
\def\beq{\begin{equation}}
\def\eeq#1{\label{#1}\end{equation}}
\def\eeqn{\end{equation}}

\def\beqa{\begin{eqnarray}}
\def\eeqa#1{\label{#1}\end{eqnarray}}
\def\eeqan{\end{eqnarray}}

\let\bar=\overbar

\def\Dslash{\not{\hbox{\kern-4pt $D$}}}
\def\dslash{\not{\hbox{\kern-2pt $\del$}}}

\def\msb{{\bar{\ssstyle M \kern -1pt S}}}

